\documentclass[epsf,preprint,12pt]{elsarticle}
\usepackage{epsfig}
\usepackage{amsmath}

\newcounter{bla}

\newcommand{\sub}[1]{\ensuremath{_{\mbox{\scriptsize \,#1}}}}
\newcommand{\supers}[1]{\ensuremath{^{\mbox{\scriptsize #1}}}}
\def\KT {$k${\ensuremath{_{\mbox{\scriptsize T}}}} }
\def\epspace {$\eta$\,-$\varphi$ space }

\begin{document}
\begin{frontmatter}

\title{FFTJet: A Package for Multiresolution Particle Jet 
Reconstruction in the Fourier Domain}

\author{I.~Volobouev}
\ead{i.volobouev@ttu.edu}
\address{Texas Tech University, Department of Physics, Box 41051,
Lubbock, Texas, USA 79409}

\begin{abstract}
This article describes the FFTJet software package designed to
perform jet reconstruction in the analysis of high energy physics
(HEP) experimental data. A~two-stage approach is adopted in which
pattern recognition is performed first, utilizing multiresolution
filtering techniques in the frequency domain. Jet energy reconstruction
follows, conditional upon the choice of
signal topology.
The method is efficient, global, collinear and infrared safe,
and allows the user to identify and avoid the event topology bifurcation points
when energy reconstruction is performed.
\end{abstract}


\end{frontmatter}


{\bf PROGRAM SUMMARY}

\begin{small}
\noindent
{\em Authors:} I.~Volobouev                                   \\
{\em Program Title:} FFTJet                                   \\
{\em Program URL:} http://projects.hepforge.org/fftjet/  \\
 {\em Journal Reference:}                                      \\
 {\em Catalogue identifier:}                                   \\
{\em Licensing provisions:} MIT License                       \\
{\em Programming language:} C++                               \\
{\em Computer:} Any computer with a modern C++ compiler       \\
{\em Operating system:} UNIX, Linux                           \\
 {\em Classification:} 11.9                                   \\
{\em External routines/libraries:} FFTW, OpenDX (optional) \\
{\em Nature of problem:} Particle jet reconstruction in 
high energy physics collider data \\
{\em Solution method:} The task is split into two distinct stages: 
pattern recognition and jet energy reconstruction. Multiresolution 
filtering techniques in the Fourier domain are utilized at the pattern 
recognition stage. Energy reconstruction is accomplished using 
weights generated by cluster membership functions. Both crisp 
and fuzzy clustering is supported.\\
{\em Running time:} Variable, depends on the
                    event occupancy and algorithm used.\\
\end{small}




\section{Introduction}

The problem of jet reconstruction is ubiquitous in the analysis of particle
physics experimental data. The celebrated asymptotic
freedom of partons at high energies~\cite{ref:asfree}
leads to a description of
particle interactions in which there is little or no interference between
the hard scattering and the hadronization stages. The kinematic
properties of parent partons are imprinted on jets, and this allows the
experimentalists to reconstruct the physics at the characteristic
scales of a few fermi (10$^{-15}$~m) from the large-scale energy
deposition structures observed in a~particle detector.

The FFTJet software package described in the current article allows
its users to
implement a variety of jet reconstruction scenarios following
the same basic two-stage approach: first, pattern recognition is
performed whereby ``preclusters'' are found in the
\epspace$\!\!$\footnote{$\eta$ and $\varphi$ are the variables
which define the direction of the energy deposit.
$\varphi$ is the azimuthal angle, while the meaning of $\eta$
is user-selectable (typically, rapidity or pseudorapidity).} and then
jet energies are reconstructed using preclusters as initial approximate jet
locations. This approach has several
important advantages over the cone and \KT jet reconstruction
algorithms~\cite{ref:snowpot} used at currently operating hadron collider experiments:
\begin{itemize}
\item The techniques used to determine jet energies are not necessarily
      optimal for determining the event topology ({\it i.e.},
      the number of jets).
      These problems are distinct and should be solved separately.
      The tasks of defining "what is a jet", locating jets, selecting
      the event topology, and reconstructing jet energies are cleanly
      separated.
\item The knowledge of the jet shape asymmetry in the \epspace can
      be effectively utilized which results in a superior algorithm
      performance in the presence of a magnetic field.
\item Provisions can be made for efficient suppression of the detector noise
      both at the pattern recognition and at the energy reconstruction stages.
\end{itemize}
The computational complexity of the pattern recognition stage is
$O(S N \ln N)$, where $N$ is on the order of the number of towers
in the detector calorimeter and $S$ is the user-selectable
number of angular resolution scales (cone and \KT algorithms use only
one resolution scale). This complexity is independent from the
detector occupancy and thus allows for predictable execution
times which can be important for online use. The computational
complexity of the jet energy reconstruction stage is $O(J M)$ where
$J$ is the number of jets found and $M$ is the number of objects
(4-vectors) used to describe the event energy flow ($M \le N$).

The main computational engine behind the pattern recognition
stage is Discrete Fast Fourier Transform (DFFT). The FFTJet
package is designed to take advantage of widespread availability
of DFFT implementations. The pattern recognition code can be easily
adapted to run on a~variety of hardware platforms including
Digital Signal Processors (DSPs) and Graphics Processing Units (GPUs).

\section{Emphasis on Pattern Recognition}

The necessity of improving pattern recognition capabilities
of jet reconstruction algorithms has been recognized
both in the context of multijet, high occupancy events and
in the cases when massive particles decaying hadronically via
electroweak interaction ($W$ and $Z$ bosons, top quarks)
are sufficiently boosted so that the energy flow of their
decay products can not be partitioned into
well-separated jets. Due to software
availability, LHC-targeted studies of particle physics
processes of this 
kind utilized predominantly sequential recombination techniques
(see Ref.~\cite{ref:jetography} for a~recent review).
The FFTJet package
is designed to provide advanced pattern recognition
performance in a~global algorithm.

The jet reconstruction model implemented in FFTJet
was originally inspired by Refs.~\cite{ref:cheng} and~\cite{ref:modetree}
and initially proposed in~\cite{ref:igv2006}.\footnote{Similar
clustering and pattern recognition
schemes have been introduced in various
disciplines, {\it e.g.,}~\cite{ref:wong, ref:roberts, ref:nakamura, ref:scalespclus}.
Early influential works include~\cite{ref:witkin, ref:koenderink, ref:lindebergThesis}.}
The study by Cheng~\cite{ref:cheng} establishes an~important connection
between the iterative cone algorithm\footnote{The iterative cone
algorithm is known as the ``mean shift''
algorithm~\cite{ref:meanshift} in the pattern recognition literature.} and
kernel density estimation
(KDE)~\cite{ref:kde}. Cheng proves that
the locations of stable cone centers correspond to
modes (peaks) of the energy
density built in the \epspace using kernel density
estimation with the Epanechnikov kernel. That is, all
such centers can be found by convolving the empirical energy density
$$
\rho\sub{emp}(\eta, \varphi) = \sum_{i} \varepsilon_i \delta^{2} (\eta - \eta_{i}, \varphi - \varphi_{i})
$$
with the function
$$
\mbox{Epanechnikov}(\eta, \varphi) = \left\{ \begin{array}{ll}
                           1 - (\varphi^2 + \eta^2)/R^2, & \varphi^2 + \eta^2 < R^2\\
                           0, & \varphi^2 + \eta^2 \ge R^2
                           \end{array}
                       \right.
$$
and then funding all local maxima of the convolution\footnote{The reader
familiar with the concept of ``Snowmass potential''~\cite{ref:snowpot} will recognize
that this potential reproduces such a~convolution up to a negative constant factor.}.
Here, $\varepsilon$ is an energy
variable whose precise meaning is user-defined (typically, transverse momentum
or transverse energy of a~particle, calorimeter tower, {\it etc.}), 
$\delta^{2} (\eta, \varphi)$ is the two-dimensional delta function,
and the sum is performed over all energy deposits expected to be
jet constituents (typically, leptons and photons produced
in the hard scattering process are excluded). $R$
is the cone radius in the \epspace$\!\!$.

\begin{figure}[h!]
\begin{center}
\leavevmode
\epsfxsize=2.0in
\epsfbox{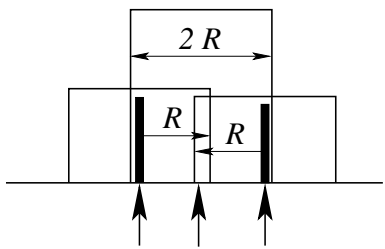}
\caption{The locations of the three stable clusters reconstructed by the ideal
iterative cone algorithm from two energy deposits are shown by the arrows
at the bottom.}
\label{split}
\end{center}
\end{figure}

The connection between the
iterative cone algorithm and KDE immediately suggests 
an~efficient implementation of a seedless cone algorithm: one should
discretize the calorimeter signals (or MC particles) on a regular
grid in the \epspace$\!\!$, perform the convolution
by DFFT, and find the peaks. This approach, however, does not
address an~important problem inherent in the cone-based jet
reconstruction. This problem manifests itself as the pattern recognition
ambiguity illustrated in Figure~\ref{split}.
Two energy deposits of similar magnitude
separated by a distance larger than $R$ but
smaller than $2 R$ produce three stable cone centers whose positions
are shown with the arrows at the bottom of the figure. In various
implementations of cone-based jet reconstruction procedures, this
problem is usually addressed by the
``split-merge'' stage which happens after the
stable cone locations are determined.\footnote{Other, less common
techniques are ``split-drop'' and ``progressive removal''~\cite{ref:jetography}.} During this stage,
jets are merged if the energy which falls into the common region
exceeds a~predefined fraction of the energy of the jet with
smaller magnitude. Even if the search for stable
cones is performed in the infrared and collinear safe manner, the
outcome of the split-merge stage is often unstable
because the decision on whether to merge the two jets depends on the
minute details of the energy deposition structure.

\begin{figure}[h!]
\begin{center}
\leavevmode
\epsfxsize=5.5in
\epsfbox{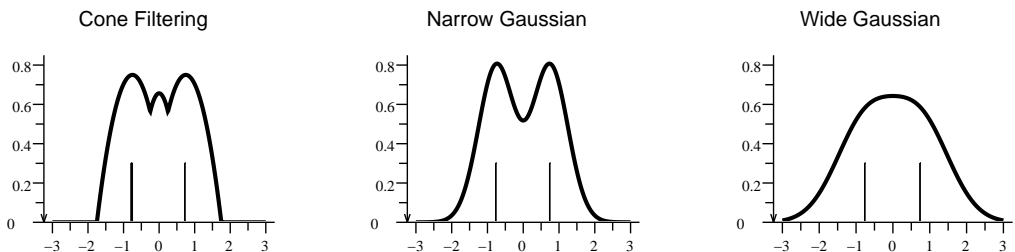}
\caption{The convolution of two energy deposits with the Epanechnikov kernel (left)
and with two Gaussian kernels of different width.}
\label{trouble}
\end{center}
\end{figure}

Another way to look at this problem and a possible solution is illustrated
in Figure~\ref{trouble}.
The third stable cone center between the two energy
deposits happens because the sum of two Epanechnikov kernels placed
at the locations of the deposits has a spurious peak in the middle. However,
there is a variety of kernels which do not suffer from this problem.
In particular, the Gaussian kernel produces either two (narrow kernel) or one
(wide kernel) peaks, as shown. Even though
one still has to address the question
of choosing the kernel width, the Gaussian kernel has
a very important advantage: the whole split-merge stage is no
longer necessary.

An intelligent choice of the kernel width (or the $R$ parameter in the \KT
and cone algorithms) can not be performed until
some assumptions are made about the expected jet shapes. In fact,
optimal choice will be different for different signals. For example,
a data analysis which searches for high energy dijet events with two well-separated
jets is likely to make very different assumptions about jets from a~data
analysis which looks for $t\bar{t}$ events in the all-hadronic, 6-jet mode.
Moreover, the optimal width is not necessarily the same for every jet
in an event, as low momentum jets tend to have wider angular profiles,
especially in the presence of magnetic field. Because of this, it is
interesting to look at the jet structure of an event using a variety of
kernel width (cone radius, {\it etc.}) choices. In the limit of continuous
kernel width we arrive to a description of event energy flow known
as ``mode tree'' in the nonparametric statistics literature or
``scale-space image representation'' in the computer vision theory.
The information contained in such a description permits multiple
optimization strategies for jet reconstruction which will be discussed
in section~\ref{sec:strategies}.

The FFTJet approach differs significantly from the majority of
jet reconstruction algorithms in the following way: 
no attempt is made to define outright what is a jet and how
particle jets should look like. Instead, the {\em package user} 
introduces an operational jet definition by selecting
the pattern recognition kernel and the jet membership
function which describes the jet shape. Given this definition, the
code efficiently searches for jet-like structures in the event
energy flow
pattern. The rationale for this view of jet reconstruction comes
from the realization that, in practice, there is no single
optimal jet definition for different particle processes.
Also, the instrumental effects
(nonlinear response and finite energy resolution of the
calorimeter, presence of the
magnetic field in the detector, material in front of
the calorimeter, pile-up, noise, 
{\it etc.}) must be taken into account,
and will almost surely dominate the systematic error
of any precision measurement based on jets.
Therefore, a~unified algorithmic definition of ``what is a jet'' 
can not be achieved across different measurements in a variety of
experimental setups.

\section{Top-Level Steps of the Algorithm}

The users of the FFTJet package are expected to reconstruct jets
using the following sequence of steps:
\begin{enumerate}[Step 1.]
\item The event energy flow is discretized using a grid in the
      \epspace$\!\!$.

\item The discretized energy distribution is convolved with a kernel function
      $K(\eta, \varphi, s)$,
      where $s$ is the resolution
      scale parameter which determines the width and,
      possibly, the shape of the kernel. Many standard kernel functions are
      included in the FFTJet package, and user-defined kernels can be
      seamlessly added as well. The convolution is performed by DFFT.

\item The peaks of the convolved energy distributions are found. These
      are potential ``preclusters''.

\item Preclusters with small magnitudes are eliminated in order
      to suppress the calorimeter noise\footnote{Even in the
      absence of such noise, fake preclusters will be detected due to
      the presence of round-off errors in the DFFT procedure.}.

\item Steps 2 through 4 are repeated as many times as necessary
      using different values of $s$. The resulting preclusters are
      arranged in the ``clustering tree'' structure.

\item Using the clustering tree information and assumptions about
      the signal spectrum, a~decision is made about the event topology
      by choosing a set of preclusters. These preclusters are passed to
      the jet energy reconstruction stage.

\item Jet constituents are determined
      as follows. The event is viewed as a~collection of energy deposits
      characterized by their direction $(\eta, \varphi)$ and
      energy variable $\varepsilon$.
      Depending on the environment in which the code is used,
      these deposits can originate from
      detector calorimeter cells, reconstructed tracks,
      Monte Carlo particles, {\it etc.}
      A~cluster membership function
      $M_{j}(\eta - \eta_j, \varphi - \varphi_j, \varepsilon, s_j)$
      is associated with
      each precluster $j$ at angular coordinates
      $(\eta_j, \varphi_j)$ and scale $s_j$. There is also a~membership function
      for the unclustered energy/underlying event. 
      The cluster membership functions
      are evaluated for every energy deposit in the event.
      In the ``crisp'' clustering scenario, an energy deposit is assigned to the jet
      whose membership function for this deposit is the largest. In the ``fuzzy''
      scenario, the deposit is split between all jets with weights
      proportional to their respective membership function values
      (the sum of all weights is normalized to 1 to ensure energy conservation).

\item Jet energies are calculated according to one of the standard
      recombination schemes using weights determined in the previous step.
\end{enumerate}
This sequence will work well for a wide variety of HEP data analyses.
Yet, if necessary, the balance between the code speed and the precision
of jet energy determination can be shifted in either direction.
For example, to speed things
up, the pattern recognition can be performed at a~single
predefined scale $s_0$. Alternatively, to further improve the jet energy
resolution, the last two steps of the algorithm can be applied iteratively.
In such a procedure,
the jet directions $(\eta_j, \varphi_j)$
and the membership function scales $s_j$ are
updated at each iteration using
reconstructed jets from the previous iteration until
some convergence criterion is
satisfied.\footnote{This iterative procedure is known as
``generalized mean shift'' or ``expectation maximization''
algorithm~\cite{ref:numrecipes, ref:equiv}.}

\section{Algorithm Details}

The FFTJet package is designed with the goals of flexibility
and extensibility in mind. The code is written in the
standard C++ programming language. Most
of the FFTJet classes are either templates or
they inherit from abstract base classes. This freedom
of choice allows the user to tailor FFTJet easily to the needs of
a~particular data analysis and software environment, but it can
also be daunting at the beginning. The intent of this section
is to provide helpful guidelines for making important decisions
about algorithm details and to explain how these details
may affect the algorithm performance.

\subsection{Energy Discretization}

The purpose of the energy discretization step
is to create a grid equidistant in the \epspace and
to populate this grid with the observed energy
values. Several typical use cases are envisioned:
\begin{itemize}
  \item The analysis is using calorimeter data collected by an experiment
        (or any other data with intrinsic granularity).
        In this case the cell sizes in $\eta$ and $\varphi$ should be
        chosen in such a way that they reflect the spacing
        of the calorimeter towers. At the same time, the number of bins should
        allow for subsequent efficient DFFT of the gridded data, so that
        exact powers of two are preferred. In case the calorimeter granularity
        is not constant throughout the full $\eta$ acceptance range or
        if the calorimeter towers are not rectangular, the FFTJet package
        includes regridding facilities which can aid in mapping the data
        onto rectangular equidistant grid with minimal loss of
        information.

  \item The event energy flow representation
        includes tracking data obtained by particle flow
        analysis or by other similar means. In this case the grid is filled
        in a manner which preserves the $\eta$\,-$\varphi$ centroid of each
        energy deposit. The grid size should be chosen in such a way that
        the binning effects do not prevent the user from seeing the smallest
        interesting detail. A good rule of thumb
        (based on the Nyquist sampling theorem) is that in each direction
        the discretization grid granularity
        should be two times finer than the typical size of such a detail.

  \item Monte Carlo particle data are analyzed, and the user 
        wants to simulate
        binning effects of a realistic calorimeter. In this case
        the FFTJet
        gridding code can function as a histogram
        with cylindrical topology.
\end{itemize}

\subsection{Pattern Recognition Kernel}
\label{sec:patker}

Pattern recognition is performed in FFTJet by convolving
the discretized event energy flow with a kernel function and then finding
the peaks of the obtained smoothed energy distribution.
Peak coordinates are determined with subcell precision: the
smoothed energy flow shape is fitted in a $3 \times 3$ rectangle
near each local maximum
with a two-dimensional quadratic polynomial by least squares method.
The location of the polynomial maximum is then used as the
peak position or
the peak is discarded if the Hessian matrix of the fitted
polynomial is
not negative definite.

The optimal choice of the pattern recognition kernel will depend on
the analysis strategy and the amount of information the user has about
the signal and the background at the time pattern
recognition is performed. The typical role which kernel plays is that
of the low-pass spatial filter in the $\eta$-$\varphi$ space: it is
supposed to enhance jet-like structures present in the event and it
has to suppress higher spatial frequency random noise present
due to fluctuations
in the showering and hadronization processes, instrumental noise, {\it etc}.
If signal and background properties are well understood, the filter
can be designed to provide optimal pattern recognition for the
process of interest (Wiener filtering~\cite{ref:numrecipes}).
This, however, is not a~typical usage for a jet clustering algorithm
in a HEP experiment. Instead, it is often more desirable to cluster
jets in a generic manner consistent with a~wide variety of
signal and background hypotheses.

The FFTJet package can aid the user in implementing several different
pattern
recognition strategies. A fast and efficient jet finding
can be performed at a~single resolution scale (which is similar to
using jets reconstructed at one cone radius). Here, a proper kernel
choice allows not only to avoid the split-merge stage
but also to take into account the nonsymmetrical jet shape in
the presence of a~magnetic field. Indeed, at sufficiently high
values of transverse momenta (above $p_{T} = 10$~GeV/$c$ or so) the width
of the transverse jet energy profile scales inversely proportional to
jet $p_{T}$. At the same time, the angular distance between the
direction of the jet axis and the location where charged particles
hit the calorimeter in the magnetic field also scales in the inverse
proportion to particle's $p_{T}.$\footnote{More precisely, 
$\sin(\Delta \varphi)$ is inversely proportional to
particle's radius of gyration and, therefore, inversely
proportional to particle's $p_{T}$ as well.} This leads to a~situation
in which the jets have a~characteristic $\eta$ to $\varphi$ width ratio
which remains stable through a~wide range of jet energies.

Modern HEP experiments often employ cone and \KT algorithms 
for jet reconstruction using several different values of the $R$
parameter which determines characteristic jet width. The FFTJet package
takes this strategy to its logical conclusion and allows
the user to view the energy flow in the event as a collection
of jet structures reconstructed using a continuous range of
angular resolution scales. In order to locate patterns which
correspond to actual physics processes in this ``scale space''
view of jet reconstruction, it becomes essential
to establish hierarchical
relationships between structures found at larger and smaller scales.
If we want to establish these relationships in a~meaningful way,
the number of jets found should decrease when the resolution
scale increases. This places an~important technical requirement
on the kernel or sequence of kernels used at different scales:
the number of peaks found after convolving the kernel with
the event energy structure should
decrease with increasing scale, no matter how the event energy
flow looks like.

It turns out that, in the form stated above, this requirement is very strict.
It is not known at this time whether such a kernel or a sequence
of kernels can actually be constructed.\footnote{Gaussian kernel
satisfies this requirement in one dimension. There are
reasons to believe that
it is impossible to satisfy this requirement in a~multidimensional
space~\cite{ref:lindebergThesis}.
However, to the author's knowledge, this impossibility has not
been strictly proven.}
Nevertheless,
the Gaussian kernel comes very close to fulfilling this requirement
for all practical purposes.\footnote{In more
than one dimension, situations in which the number
of peaks increases with increasing scale do arise, albeit infrequently.
For example, three 
energy deposits of equal magnitude placed at the corners
of an equilateral triangle will,
for a certain narrow range of resolution scales,
produce a spurious fourth peak at the triangle center~\cite{ref:ngauss}.}
In general, an optimal choice of a pattern recognition kernel should
result both in good local properties of the reconstructed jets
(robustness with respect to small variations in jet energy flow and
resistance to noise) and in good scaling properties: the
event topology should vary naturally in the scale space.

Perhaps, the most useful general-purpose multiresolution
kernel implemented in the
FFTJet package is the Gaussian
kernel corrected for the energy flow discretization
effects.
This kernel is the Green's function of the two-dimensional
anisotropic diffusion equation with
the discretized Laplacian operator (the rationale for
this approach and the formula for the isotropic diffusion case
is given in \cite{ref:discretescalespace}).
Unlike the standard Gaussian kernel which 
is strongly affected by binning effects when its width
becomes comparable to the grid bin size,
the corrected kernel behaves meaningfully at small scales, and
gracefully converges to the discrete delta function
at the zero scale limit. The kernel is defined by its
Fourier transform representation:
\[
\begin{array}{rcl}
\mbox{Re}(F(u, v)) & = & \exp \left(\frac{\sigma_{\eta}^2}{(\Delta \eta)^2} (\cos (u) -1) + \frac{\sigma_{\varphi}^2}{(\Delta \varphi)^2} (\cos (v) -1) \right), \\
\mbox{Im}(F(u, v)) & = & 0,
\end{array}
\]
where

$u = \frac{2 \pi k}{N_{\eta}}$, $k \in \{0, 1, ..., N_{\eta}-1\}\,$ is the $\eta$ frequency.

$v = \frac{2 \pi m}{N_{\varphi}}$, $m \in \{0, 1, ..., N_{\varphi}-1\}\,$  is the $\varphi$ frequency.

$\Delta \eta = \frac{2 \pi}{N_{\eta}}$ is the {\it effective} width of the grid cells in $\eta$ (scaled so that the full $\eta$ range of the grid is $2 \pi$).

$\Delta \varphi = \frac{2 \pi}{N_{\varphi}}$ is the width of the grid cells in $\varphi$.

$\sigma_{\eta}$ is the {\it effective} kernel width parameter in $\eta$. In the
limit of small cell sizes and when $\sigma_{\eta} \ll 2 \pi$, it corresponds
to the standard deviation of the Gaussian kernel.

$\sigma_{\varphi}$ is the kernel width parameter in $\varphi$.

In addition to its excellent performance in the multiresolution context,
the Gaussian kernel has another useful feature.
For well-separated, symmetric jets
the peak magnitude dependence on the resolution
scale, $m(s)$, is the Laplace transform of the transverse
energy profile.\footnote{A proper
selection of the kernel $\eta$ to $\varphi$ width ratio
makes it a good approximation even in the presence of a strong
magnetic field.} Therefore, a reasonable estimate
of the jet transverse energy can be obtained from
$$E_{T,0} = A \,\lim_{s \rightarrow \infty} s^2 m(s),$$
where $A$ is a proper normalization
constant which depends on the binning of the energy discretization
grid.\footnote{
In the actual code which evaluates the limit
one can exchange the parameter $s$ with the parameter $\alpha = s^p,
\,p < 0$, and then extrapolate towards $\alpha = 0$.} This initial
estimate can be used by
the energy recombination stage of the algorithm.

\subsection{The Clustering Tree}

The clustering tree represents the agglomeration of peaks (preclusters) found
by multiresolution spatial filtering into a single hierarchical structure.
The tree is constructed using a distance function.
A precluster found at some resolution scale $s_i$ is assigned
a parent from the previous (larger) resolution scale $s_{i-1}$ as follows:
the distance between the precluster at the scale $s_i$ is
calculated to all preclusters at the scale $s_{i-1}$.
The precluster at the scale $s_{i-1}$ with the smallest such
distance becomes the parent. This simple agglomeration strategy
follows the approach of Ref.~\cite{ref:modetree} and has the
advantage that the obtained tree structure can also be 
utilized as a balltree~\cite{ref:balltree}. Other agglomeration
strategies are possible~(see \cite{ref:scalespclus} and references therein)
and may be implemented in the future FFTJet releases.

The choice of the function which defines the distance between
the preclusters is up to the user of the 
package. The implementation
must at least ensure that the
distance can never be negative, the distance from any precluster
to itself is zero, the distance is symmetric for preclusters found at the same
resolution scale, and that the triangle inequality
is satisfied for any three
preclusters.\footnote{That is, for each resolution scale
precluster variables must form a pseudometric space. For different
scales, the commutativity requirement of the distance function
can be dropped because the preclusters are naturally ordered by scale.
}
The package
itself provides one such distance function defined as
$d = \sqrt{\left(\frac{\Delta \varphi}{h_{\varphi}}\right)^2 + 
\left(\frac{\Delta \eta}{h_{\eta}}\right)^2}$, independent from
peak magnitudes and
resolution scales used. The bandwidth values $h_{\eta}$ and
$h_{\varphi}$ are typically chosen so that $h_{\eta} / h_{\varphi} = r$
 and $h_{\eta} h_{\varphi} = 1$, and then $r$ is the only parameter
needed to define the distance function.

Once the parent/daughter relationships are established between
preclusters found at different resolution scales, dependence of various precluster
characteristics on the scale parameter can be analyzed.
By default, FFTJet calculates the following precluster properties:
\begin{itemize}
\item The speed with which the peak magnitude changes as the function of scale. This is an approximate value of $\frac{d \log(m(s))}{d \log(s)}$.
\item The speed with which the precluster location drifts in the scale space. If the distance between precluster is defined by the angular distance $d$ described above,
this becomes $\frac{|d\,\vec{r}\,|}{d \log(s)}$, with
$\vec{r} = (\frac{\varphi}{h_{\varphi}}, \frac{\eta}{h_{\eta}})$.
\item Precluster lifetime in the scale space. It is computed as
$\log(s_{max}) - \log(s_{min})$
where $s_{max}$ and $s_{min}$ define
the range of resolution scales for which the
precluster exists as a~distinct feature of the energy
distribution. Typically, the lifetime is traced
from the smallest scale in the clustering tree
to the scale where the precluster becomes
a part of a~larger precluster. If the tree is
constructed using a pattern recognition kernel which generates
spurious preclusters, this quantity can be used for trimming
such preclusters.
\item Distance to the nearest neighbor precluster at the same resolution scale.
\end{itemize}
Together with the precluster locations, scales, and peak magnitudes, these quantities
are collected in a single class which describes precluster
properties in FFTJet. Each node of the clustering tree is associated
with one object of this class.

\begin{figure}[h!]
\begin{center}
\leavevmode
\epsfxsize=5.3in
\epsffile{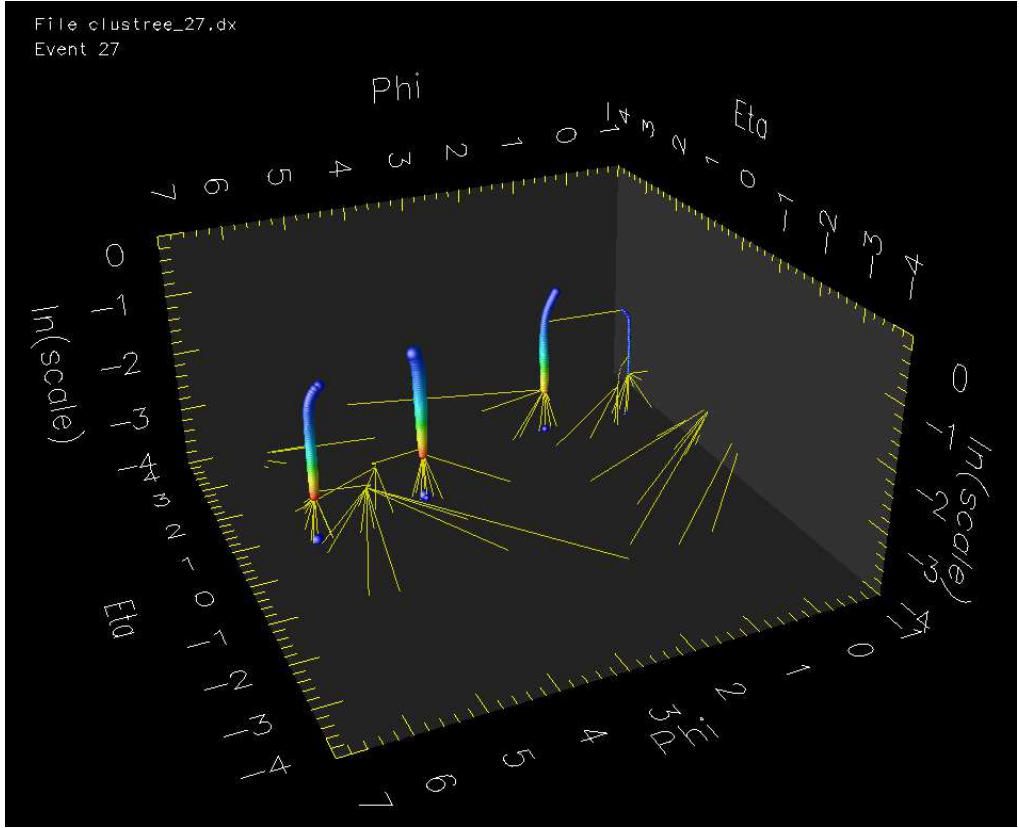}
\caption{An example clustering tree image generated by OpenDX for
a four-jet event. Here, the quantity
$s^2 m(s)$, where $m(s)$ is the precluster magnitude, is mapped into
the glyph size and the scale-normalized Hessian
blob detector~\cite{ref:blobdetect} is mapped into the glyph color.
The $\varphi$ variable
wraps around so that 0 and $2 \pi$ correspond to the same location. This is
why you see several connections apparently ending at $\varphi = 0$: they
actually ``tunnel'' from the right side of the image to the left and
continue towards the cluster near $\varphi = 2 \pi$.}
\label{opendxexample}
\end{center}
\end{figure}

The FFTJet package contains facilities for visualizing clustering
trees with the OpenDX scientific visualization system~\cite{ref:opendx}.
Various precluster properties can be mapped into the size and color
of OpenDX glyphs, while the precluster location in the \epspace and
the precluster resolution scale are mapped into the
glyph coordinates in a three-dimensional scene.
An~OpenDX view of a~clustering
tree is shown in Figure~\ref{opendxexample}. On a computer screen,
this view can be interactively shifted, scaled, and rotated
with a virtual trackball.

\subsection{Choosing the Event Topology}
\label{sec:strategies}

To determine the event topology, the user must introduce
some assumptions about the signal properties.
The clustering tree
functionality allows for an efficient implementation
of a~variety of pattern recognition strategies tuned to
locate precluster patterns consistent with the properties
of the expected signal. A~few possible strategies are listed below.
\begin{itemize}
\item The traditional approach consists in choosing
      the single best resolution scale according to some
      optimization criterion. For example, the fraction
      of events in which the number of reconstructed jets equals
      the number of partons produced at the leading order perturbation
      theory is maximized for the signal of interest.
      In the multiscale reconstruction paradigm, this approach
      can be improved upon by avoiding situations in which the chosen resolution
      scale is close to a bifurcation point --- the scale at which
      two smaller preclusters form a bigger one.
      Near the bifurcation point the locations of the affected preclusters
      become very sensitive to small changes in the event energy
      flow.\footnote{Bifurcation points are present
      in every jet reconstruction algorithm, including those
      which are normally considered to be infrared and collinear safe.
      Most algorithms do not have the capability to detect these points.
      A more detailed discussion (albeit using different terminology)
      can be found in~\cite{ref:ojf}
      where bifurcation points manifest themselves as multiple and/or shallow
      minima in the optimization of the jet configuration.}
      This problem results in an increased uncertainty of jet energy and
      direction determination. The bifurcation points can
      be avoided by detecting them in the scale space
      with the clustering tree, and by using slightly
      modified resolution scales in case such points are found.
\item For each event, one can choose a scale for
      which the number of clusters, $J$,
      corresponds to the number of jets expected in the signal.
      In order to make sure that this jet configuration indeed represents
      a salient feature of the event energy flow, some
      characterization of the configuration stability must be provided.
      This stability can be described by the configuration ``lifetime''
      in the scale space. A reasonable lifetime function is 
      $\log(s\sub{max}(J)/s\sub{min}(J))$, where $s\sub{max}(J)$ and
      $s\sub{min}(J)$ are, respectively,
      the maximum and the minimum resolution scales for which
      the tree has exactly $J$ clusters.\footnote{If you are interested
      in comparing jet configurations with different values of $J$ then
      the appropriate lifetime function should also depend on $J$.
      For example, $J^{\alpha} \log(s\sub{max}(J)/s\sub{min}(J))$ could
      be a good choice, with $\alpha$ chosen empirically depending
      on the process under study ($0 < \alpha < 1$). More 
      complicated, powerful definition
      of the lifetime in the scale space is proposed in~\cite{ref:lifetime}.}
\item Scale-space differential blob detectors~\cite{ref:blobdetect}
      can potentially be used to identify jets.
\item Nontrivial clustering patterns are identified in the signal,
      and similar patterns are searched for in the clustering tree. For example,
      boosted resonances, such as $W$ bosons or top quarks,
      are expected to produce one wide jet at higher resolution scales
      which has a~prominent substructure at lower scales.
\item The scale is chosen separately for each jet, in a manner
      consistent with the expected event topology. For example, if the cluster
      does not split across a range of scales and its position in the \epspace
      remains stable, it will be advantageous to use a~$p_T$-dependent
      jet shape model for energy determination.
\end{itemize}

\subsection{Membership Functions}

The name ``membership function'' is borrowed from the fuzzy sets
theory~\cite{ref:fuzzysets}.
The FFTJet membership functions serve a similar purpose: they reflect
the probabilities of energy deposits to belong to jets.
However, the range of FFTJet membership function values is not limited to the
interval [0, 1], instead any non-negative real value is allowed.
The jet membership functions employed by FFTJet have a continuous
scale parameter which will be called ``recombination scale'' for
the remainder of this paper.

A membership function is associated with each precluster $j$ located
at angular coordinates $(\eta_j, \varphi_j)$: $M_{j}(\eta - \eta_j, \varphi - \varphi_j, \varepsilon, s_j)$.
Here, $\varepsilon$ is the magnitude of the transverse energy (or momentum) 
of the energy deposit located at 
angular coordinates $(\eta, \varphi)$. The recombination scale $s_j$ 
may or may not coincide with the precluster resolution scale.
Defined in this manner, the membership function is invariant with respect
to shifts in the $\eta$ and $\varphi$ coordinates but not with
respect to changing the recombination scale or permuting the jets.
The noise/unclustered energy membership function
$U(\eta, \varphi, \varepsilon)$ has no characteristic scale.

Two recombination modes are supported by the FFTJet code: ``crisp'' and ``fuzzy''.\footnote{Use of fuzzy clustering
for jet reconstruction
has been advocated earlier in~\cite{ref:ojf, ref:ojf_imp}.}
In the ``crisp'' mode,  each energy deposit is assigned to the jet
(or noise/unclustered energy) whose membership function value evaluated for that deposit
is the highest. In the ``fuzzy'' mode, each energy deposit is distributed among
all jets and the unclustered energy with weights calculated for jet number $j$ as
$$
w_j(\eta, \varphi) = \frac{M_{j}(\eta - \eta_j, \varphi - \varphi_j, \varepsilon, s_j)}{U(\eta, \varphi, \varepsilon) + \sum_{k} M_{k}(\eta - \eta_k, \varphi - \varphi_k, \varepsilon, s_k)}
$$
and for the noise/unclustered energy as
$$
w_u(\eta, \varphi) = \frac{U(\eta, \varphi, \varepsilon)}{U(\eta, \varphi, \varepsilon) + \sum_{k} M_{k}(\eta - \eta_k, \varphi - \varphi_k, \varepsilon, s_k)}.
$$
The weights calculated in this manner are normalized~by
$$
w_u(\eta, \varphi) + \sum_{k} w_k(\eta, \varphi) = 1
$$
for each $\eta$, $\varphi$ which ensures that energy and momentum
can be conserved during the recombination 
procedure.

The choice of the jet membership function and the recombination mode
is up to the user of the package. It is expected that the most precise
determination of jet energies will be achieved by using detailed
jet shape models which will be called
``detector-level jet fragmentation functions''.
Such jet models are defined by
$$
M_j(\eta, \varphi, \varepsilon, s) = \left< \frac{\partial^3 N(p_T)}{\partial \eta \,\partial \varphi \,\partial \varepsilon} \right>
$$
where $N$ is the number of energy discretization grid cells
into which a jet deposits its energy, $p_T$ is the
actual jet $p_T$, the jet direction is shifted to
$(\eta_j, \varphi_j) = (0, 0)$, and angular brackets stand for averaging over
a~large number of jets. It is natural in this case to set the recombination scale $s$
to $1/p_T$. The functions $M_j(\eta, \varphi, \varepsilon, s)$
defined in this manner are normalized~by
$$
\int M_j(\eta, \varphi, \varepsilon, s) d \eta d \varphi d \varepsilon = N(p_T)
$$
and
$$
\int \varepsilon M_j(\eta, \varphi, \varepsilon, s) d \eta d \varphi d \varepsilon = E_T \ (\mbox{or } p_T).
$$
It is unlikely that in practice one will be able to represent
these jet models by simple parametrized functional expressions.
FFTJet provides a solution to this problem in the form of
multidimensional interpolation tables.
Construction and serialization of such tables is discussed in
the FFTJet package user manual~\cite{ref:packagesite}.

Within the FFTJet framework it is possible to associate different jet membership
functions with different preclusters, so the user can take advantage of
even more detailed jet models which can depend, for example, on the
assumed jet flavor,
electromagnetic energy fraction, separation from other jets, {\it etc.}
On the other hand, simpler models will
be less susceptible to systematic errors and
model misspecifications, and can potentially
result in simplified calibration procedures. For example, 
the recombination behavior of the cone algorithm can
be reproduced within FFTJet  by using crisp clustering with the
Epanechnikov kernel used as the jet membership function.
Note that in this case the membership function is not unique: in the crisp mode
identical jets will be generated by any membership function
which depends only on $r = \sqrt{\varphi^2 + \eta^2}$ and which
decreases monotonically from a positive value when $r = 0$ to zero 
when $r = R$. For use with the cone-like algorithm, it is sufficient to specify
$U(\eta, \varphi, \varepsilon) = \epsilon$, where $\epsilon$ is a
very small positive constant. 

Within FFTJet, energy resolution performance
of the cone algorithm\footnote{The pattern
recognition performance of the cone algorithm can be
reproduced exactly by using the Epanechnikov kernel at the pattern
recognition stage. Of course,
in practice you will want to make better pattern recognition kernel choices.}
can be easily improved upon
in two ways: by introducing different bandwidth
values for $\eta$ and $\varphi$ variables
(as illustrated in Section~\ref{sec:perf} of this paper), and by
choosing the $R$ parameter separately for each jet,
in a manner consistent with the event topology discovered
during the pattern recognition stage. It may also
be interesting to apply cone-like recombination stage iteratively, using
a~procedure in which the
cone radius for the next iteration depends on the jet $p_T$
determined during the previous iteration.\footnote{For example,
one can choose $R \propto p_T^{\alpha}, \, \alpha < 0$ ($\alpha$ must be negative
to ensure convergence). The optimal
choice of $\alpha$ will depend on the balance of uncertainties
due to event occupancy, calorimeter noise and energy resolution, pileup,
out-of-cone leakage, {\it etc.} The simple choice of $\alpha = -1$ can be
advocated on the basis of kinematic arguments alone~\cite{ref:varcone}. }

\subsection{Jet Energy Recombination Schemes}

At the time of this writing, three energy recombination schemes are
supported by FFTJet code. 
The first two are straightforward weighted modifications of the schemes commonly
employed at the hadron
collider experiments. The third one sets the jet direction
to the precluster direction. The latter definition
can potentially be useful for high occupancy/high noise events or
when pattern recognition is performed with filters
optimized for some specific signal and background processes.
\begin{enumerate}[Scheme 1.]
\item Weighted 4-vector recombination scheme (often called $E$-scheme):
$$
P_j = \sum_{\eta, \, \varphi} w_j(\eta, \varphi) P(\eta, \varphi),
$$
where $P(\eta, \varphi)$ is the 4-vector associated with the 
energy deposit at $(\eta, \varphi)$.
For ``crisp'' clustering, all weights $w_j(\eta, \varphi)$ for
jet number $j$ are either 0 or 1.

\item Weighted Original Snowmass scheme (also called $E_{T}$ or $p_T$ centroid scheme):
$$
\varepsilon_j = \sum_{\eta, \, \varphi} w_j(\eta, \varphi) \varepsilon(\eta, \varphi),\ \ \ \eta_j = \frac{\sum_{\eta, \, \varphi} w_j(\eta, \varphi) \,\varepsilon(\eta, \varphi) \eta}{\varepsilon_j},
$$
$$
\varphi_j = \varphi\sub{precluster,$j$} + \frac{\sum_{\eta, \, \varphi} w_j(\eta, \varphi) \,\varepsilon(\eta, \varphi) \Delta \varphi_j}{\varepsilon_j}.
$$
The variable $\varepsilon$ is chosen by the user. Normally,
this should be either $E_T$ (if $\eta$ represents pseudorapidity) 
or $p_T$ (if $\eta$ represents rapidity).
$\Delta \varphi_j$ is defined as $\varphi - \varphi\sub{precluster,$j$}$
moved to the interval from $-\pi$ to $\pi$. 
This recombination scheme can potentially outperform the 4-vector scheme
when jets are reconstructed using calorimeter towers, and a strong magnetic
field is present in the detector.

\item Precluster direction scheme:
$$
\varepsilon_j = \sum_{\eta, \, \varphi} w_j(\eta, \varphi) \varepsilon(\eta, \varphi),\ \ \eta_j = \eta\sub{precluster,$j$},\ \ \varphi_j = \varphi\sub{precluster,$j$}
$$
\end{enumerate}
%
The energy recombination step can be performed using as input either
a~collection of 4-vectors or the discretized energy flow in the \epspace\!\!.
The latter approach may be convenient for reconstructing jets from
calorimeter data.

\subsection{Implementation}

The details of the algorithm mapping into C++ classes and
the user API are
described in the FFTJet package user manual distributed together
with the code~\cite{ref:packagesite}. The package comes with several top-level
driver classes which combine multiple steps of the algorithm into
convenient API units. Example executables are provided.
These examples illustrate package
usage with both single-scale and multiresolution
pattern recognition stages.

\section{Performance}
\label{sec:perf}

This section illustrates FFTJet performance with a cone-like,
easy to calibrate jet model and a simple synthetic dataset.
It is assumed that jets are reconstructed with a projective geometry
calorimeter placed in a 3.8~T magnetic field. The width of
each calorimeter tower in $\eta$ and $\varphi$ is taken to be
$2 \pi/64 \approx 0.10$.
Two distinct, independent
light quark jets per event are generated using PYTHIA~6.4~\cite{ref:pythia}
single jet gun, using default PYTHIA tune.
The $p_T$ of the first jet is fixed at 50 GeV/$c$ and its
direction is randomized across the face of one of the calorimeter
towers near $\eta = 0$. The second jet is directed inside a circle around
the first jet in such a manner that the distribution of 
$\Delta r = \sqrt{(\Delta \eta)^2 + (\Delta \varphi)^2}$ is flat
and the orientation of the $(\Delta \eta, \Delta \varphi)$
vector is uniform and random.
The $p_T$ spectrum of the second jet is flat between 1~and 100~GeV/$c$.
The 4-momenta of these jets are defined
as the sum of the 4-momenta of all stable particles within the jet except
neutrinos.

Stable charged particles are propagated along helical trajectories
to the assumed
calorimeter radius of 2.0~m, while the neutral particles
fly straight. The energies are accumulated for each
tower separately. Neutrinos and the charged particles which do not
have enough transverse momentum to reach the calorimeter
do not contribute. The shower development in the calorimeter
is not modeled. The calorimeter is assumed to have ideal energy
response and a~noise of $\sigma_n = 0.15$~GeV per tower. The
tower threshold of $2.5\,\sigma_n$ is applied before the
jet reconstruction is performed. In this virtual setup,
the calorimeter granularity,
noise, magnetic field, and tower threshold are not far from
the values actually used in real particle detectors, while
the ideal energy response permits characterization of the jet energy resolution
performance of different algorithms without having to disentangle
the detector effects.

The jet reconstruction algorithms compared are:
\begin{itemize}
\item {\it FFTJet-1}: the FFTJet package is configured to use a single-scale
        pattern recognition stage.
        The pattern recognition kernel is the Gaussian
        kernel corrected for the energy flow discretization effects, as described
        in section~\ref{sec:patker}. The parameter $\sigma_{\eta}$
        is set to 0.1, $\sigma_{\varphi}$ is set to 0.3, the peak magnitude cutoff
        is 0.4. Crisp clustering is used, with jet membership function represented
        by the elliptical cone:
        $$
        M(\eta, \varphi, \varepsilon, s) = \left\{ \begin{array}{ll}
          1 - \sqrt{\left(\frac{\eta}{R_\eta}\right)^2 + \left(\frac{\varphi}{R_\varphi}\right)^2}, & \left(\frac{\eta}{R_\eta}\right)^2 + \left(\frac{\varphi}{R_\varphi}\right)^2 < 1\\
          0, & \left(\frac{\eta}{R_\eta}\right)^2 + \left(\frac{\varphi}{R_\varphi}\right)^2 \ge 1
        \end{array}
        \right.
        $$
        with $R_\eta = 0.2887$ and $R_\varphi = 3 R_\eta$.
        The area of the base of such a~cone is the same as the area
        of the circle with $R = 0.5$. In this configuration, the jet
        membership function has
        no dependence on $\varepsilon$, $s$, or jet number.
        The background membership function is a small 
        constant.

\item {\it SISCone}: the SISCone algorithm~\cite{ref:siscone}
       is configured with $R = 0.5$, the minimal $p_T$ for protojets 0.5~GeV/$c$,
       and unlimited number of passes. The overlap parameter for
       the split-merge stage is set to 0.75. The standalone algorithm implementation
       is used~\cite{ref:sisconeimp}.

\item \KT$\!$: the \KT algorithm is used with $R = 0.5$ and
      minimal $p_T$ for protojets 0.5~GeV/$c$.

\item {\it Anti}-\KT$\!$: the anti-\KT algorithm~\cite{ref:antikt} is used with $R = 0.5$ and
      minimal $p_T$ for protojets 0.5~GeV/$c$. Both \KT and anti-\KT
      algorithms implementations are taken from the FastJet package~\cite{ref:fastjet}.
\end{itemize}
The 4-vector recombination scheme is used with all algorithms. Similar
jet size parameters are chosen so that the conclusions of this study
are not expected to change with the inclusion of uncertainty contributions
from the calorimeter energy resolution and pileup.

\begin{figure}[h!]
\begin{center}$
\begin{array}{cc}
\epsfxsize=2in
\epsffile{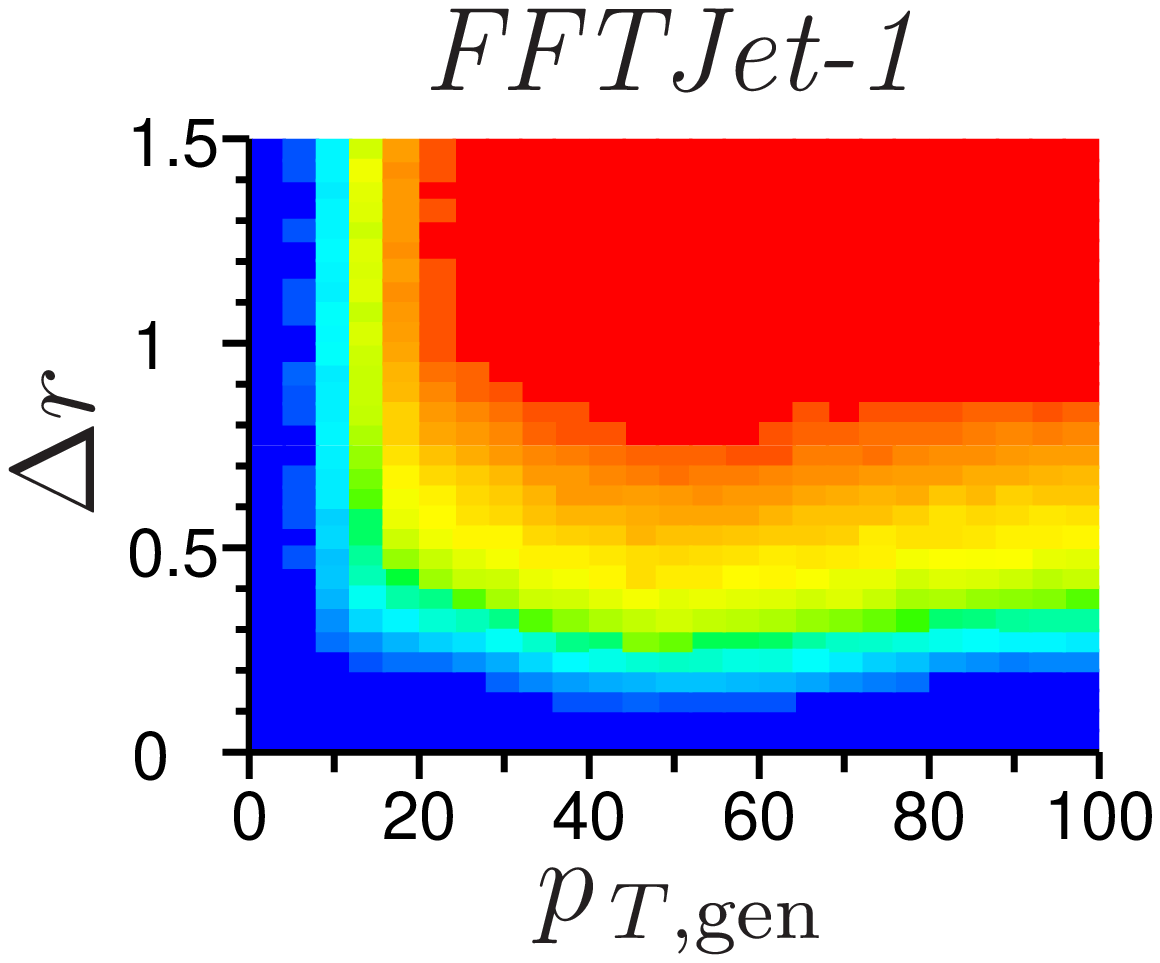} &
\epsfxsize=2in
\epsffile{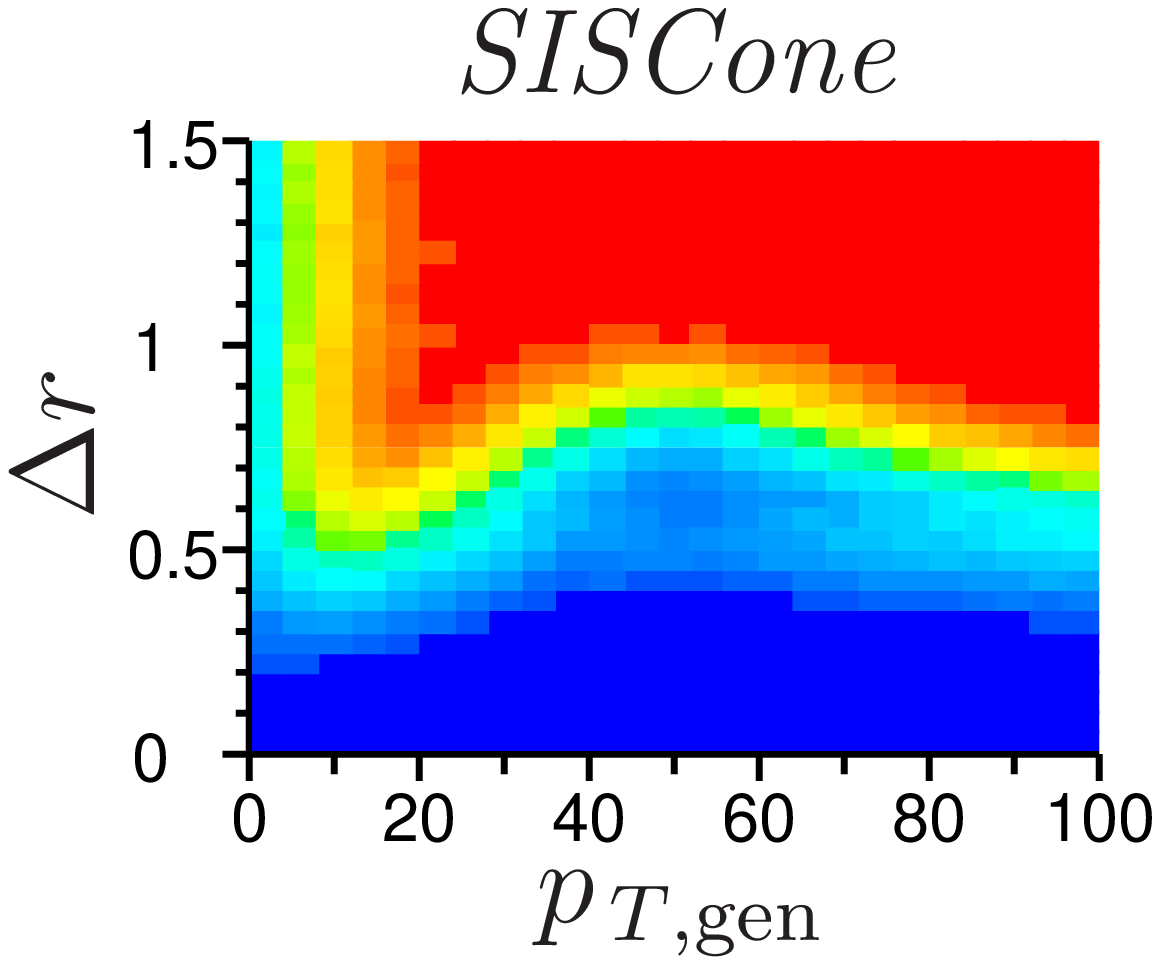} \\
\epsfxsize=2in
\epsffile{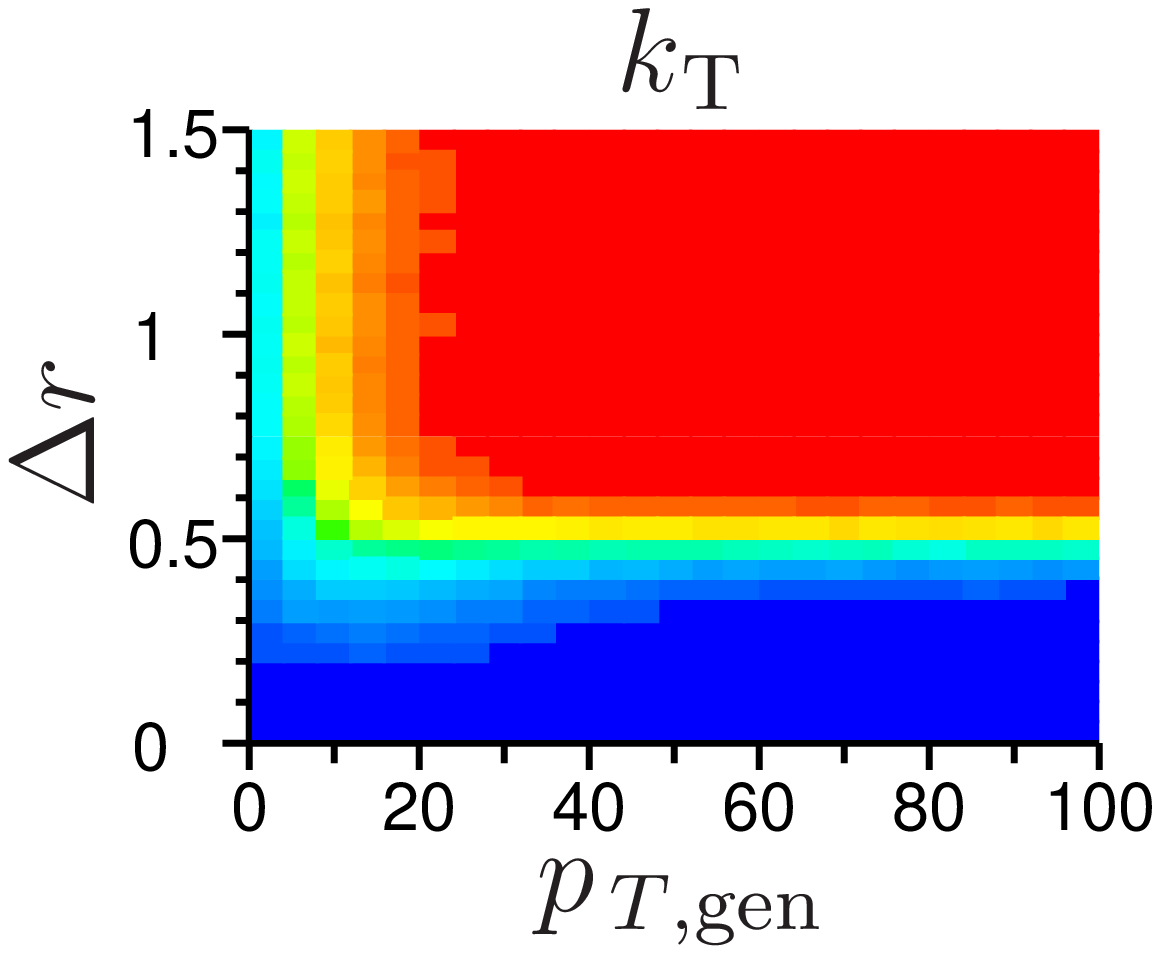} &
\epsfxsize=2in
\epsffile{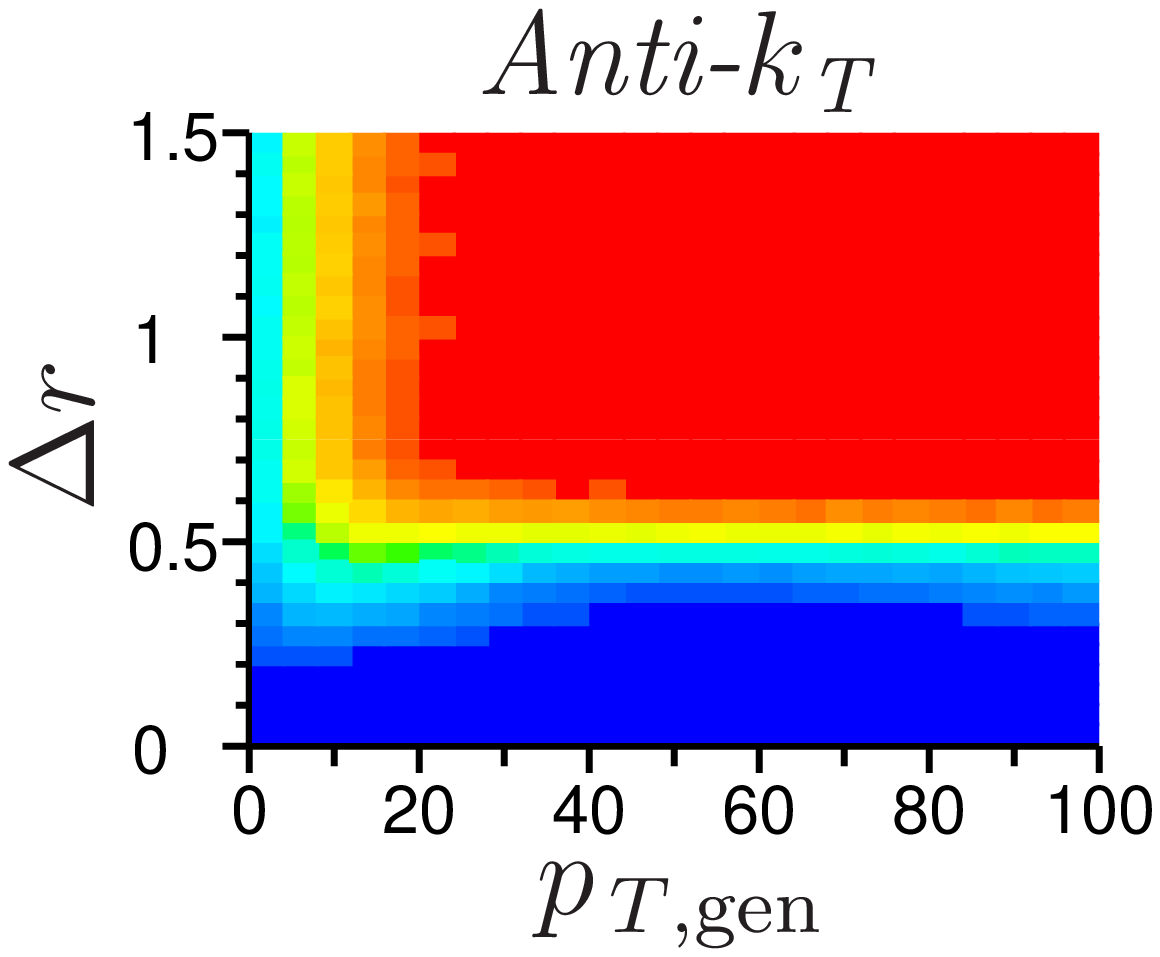} \\
\multicolumn{2}{c}{\epsfxsize=2.5in \epsffile{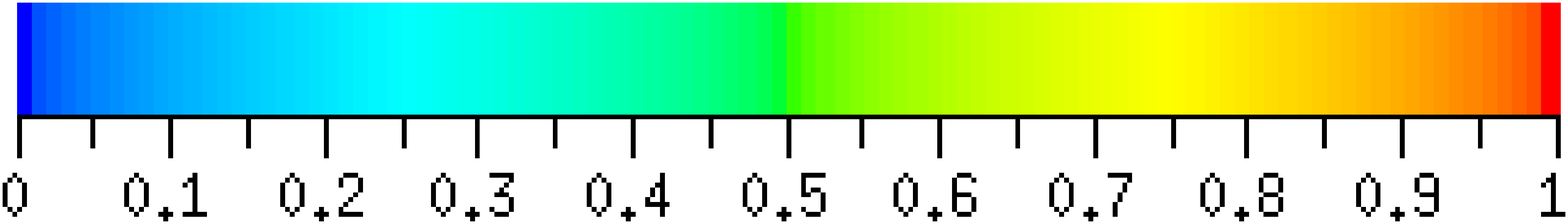}} \\
\multicolumn{2}{c}{\mbox{Efficiency color scale}}
\end{array}$
\end{center}
\caption{Jet reconstruction efficiency for different algorithms.}
\label{fig:eff}
\end{figure}

\begin{figure}[h!]
\begin{center}$
\begin{array}{cc}
\epsfxsize=1.7in
\epsffile{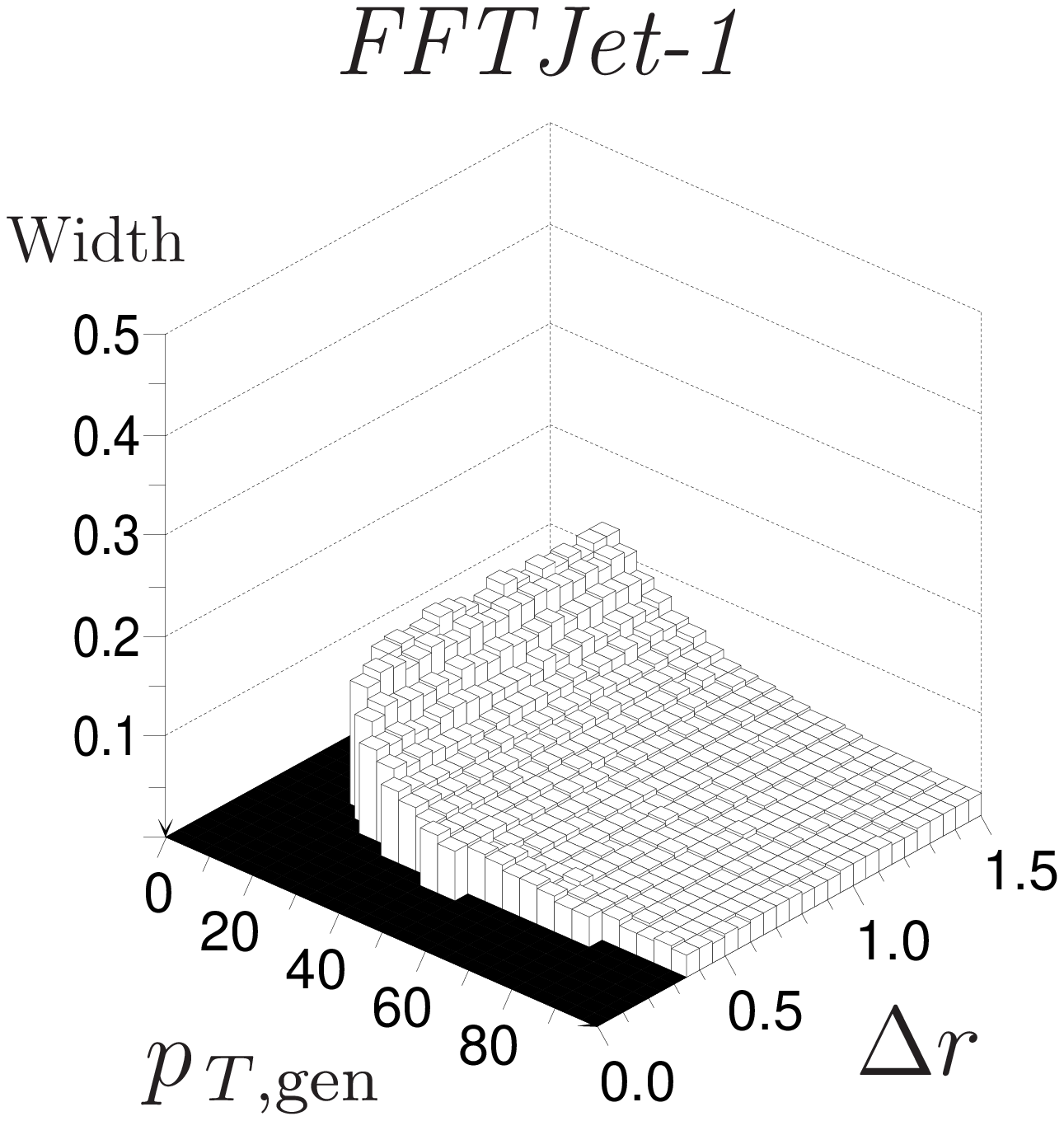} &
\epsfxsize=1.7in
\epsffile{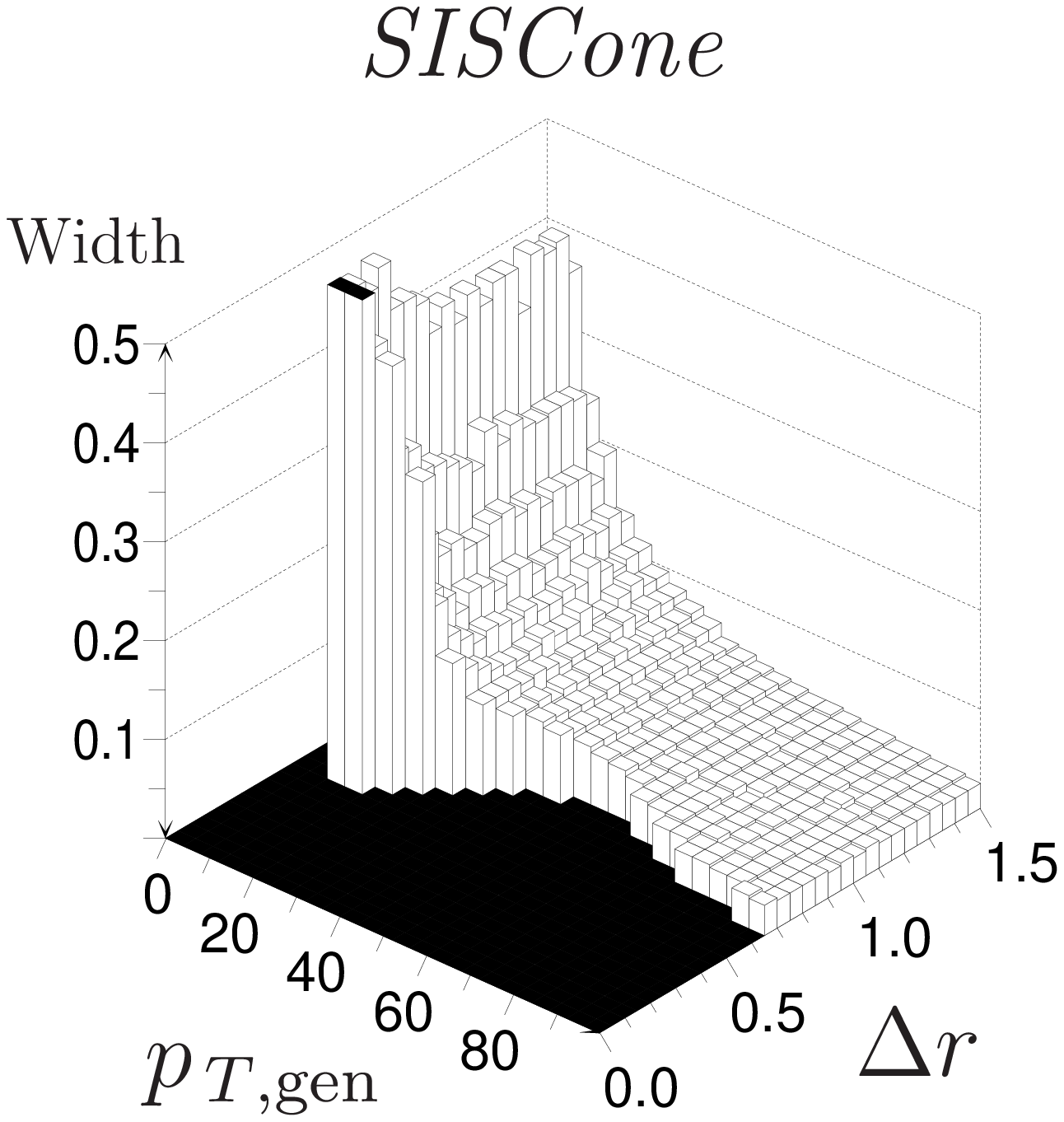} \\
\epsfxsize=1.7in
\epsffile{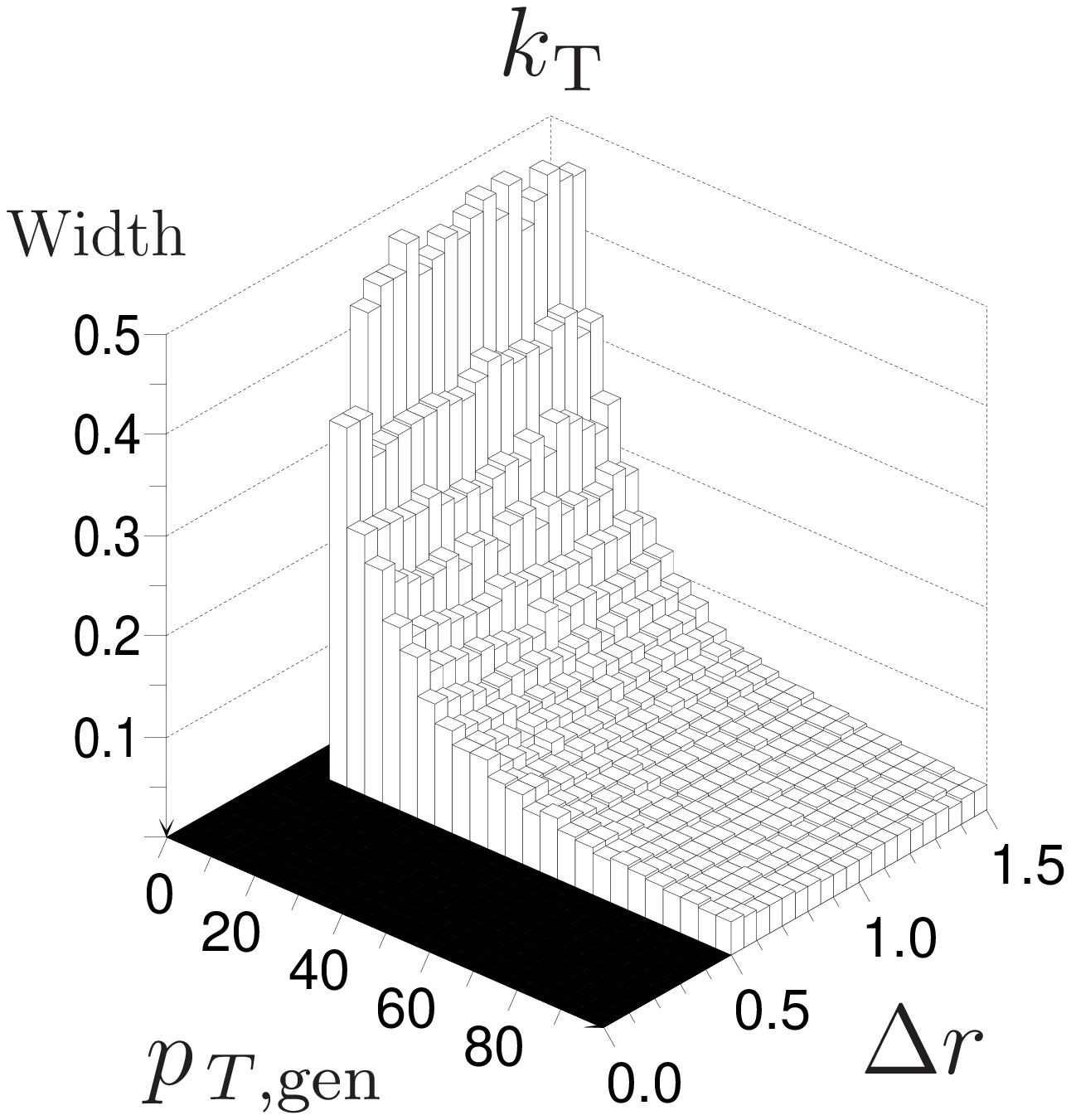} &
\epsfxsize=1.7in
\epsffile{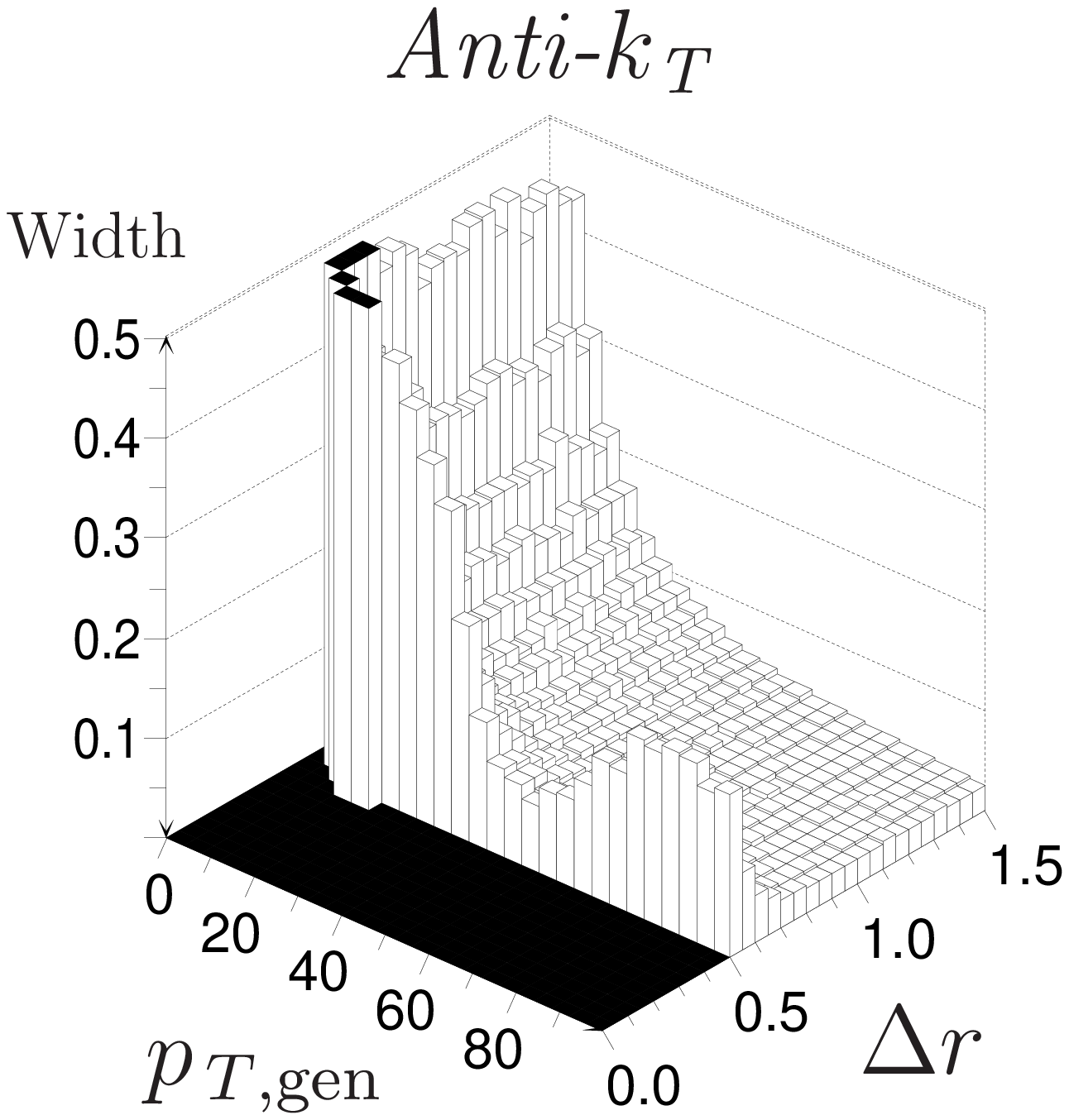}
\end{array}$
\end{center}
\caption{Relative $p_T$ reconstruction uncertainty for different algorithms.}
\label{fig:uncert}
\end{figure}

Figures~\ref{fig:eff} and~\ref{fig:uncert}
compare the event reconstruction efficiencies and the relative
$p_T$ reconstruction uncertainties for these algorithms. These
characteristics are presented as functions of $\Delta r$ between
the two generated jets and the generated transverse momentum of the
second jet, $p\sub{$T$,gen}$ (flat spectra in $\Delta r$ and $p\sub{$T$,gen}$
were chosen precisely in order to simplify construction and interpretation of these plots).
On average, $1.1 \times 10^3$ events were generated per each $p\sub{$T$,gen}, \Delta r$
bin plotted.
To determine the efficiency, the
reconstructed jets are matched to the generated jets in the \epspace$\!\!$. First, the
pair of jets with the smallest value of
$\Delta R_{ij} = \sqrt{(\eta\sub{reco,$i$} - \eta\sub{gen,$j$})^2 + (\varphi\sub{reco,$i$} - \varphi\sub{gen,$j$})^2}$ is determined
and  removed from subsequent consideration. Then the best match is determined for the
remaining generated jet. The efficiency is defined as the fraction of events
in which both $\Delta R_{ij}$ values are below 0.3.
The relative $p_T$ reconstruction uncertainty is defined as the
width of the $p\sub{$T$,reco}/p\sub{$T$,gen}$ distribution 
in each $p\sub{$T$,gen}, \Delta r$ bin (using only events which satisfy the $\Delta R_{ij} < 0.3$ requirement), where $p\sub{$T$,reco}$ is the transverse momentum of the
reconstructed jet matched to the second generated jet.
The width is calculated as one half of the difference between 
84.13\supers{th}  and 15.87\supers{th}
percentiles. For the Gaussian distribution, this robust estimate
of the width coincides with the standard deviation.
The distributions are corrected in each bin
so that the median is exactly 1
({\it i.e.}, $p\sub{$T$,reco}$ values are multiplied by a constant calculated separately for each bin).
In the plots,
the width estimate is shown only for the bins for which the reconstruction
efficiency defined in the above manner exceeds 50\% (at lower
efficiencies the width is dominated by mismatched jets). 

\begin{figure}[h!]
\begin{center}$
\begin{array}{rcl}
\epsfxsize=1.7in
\epsffile{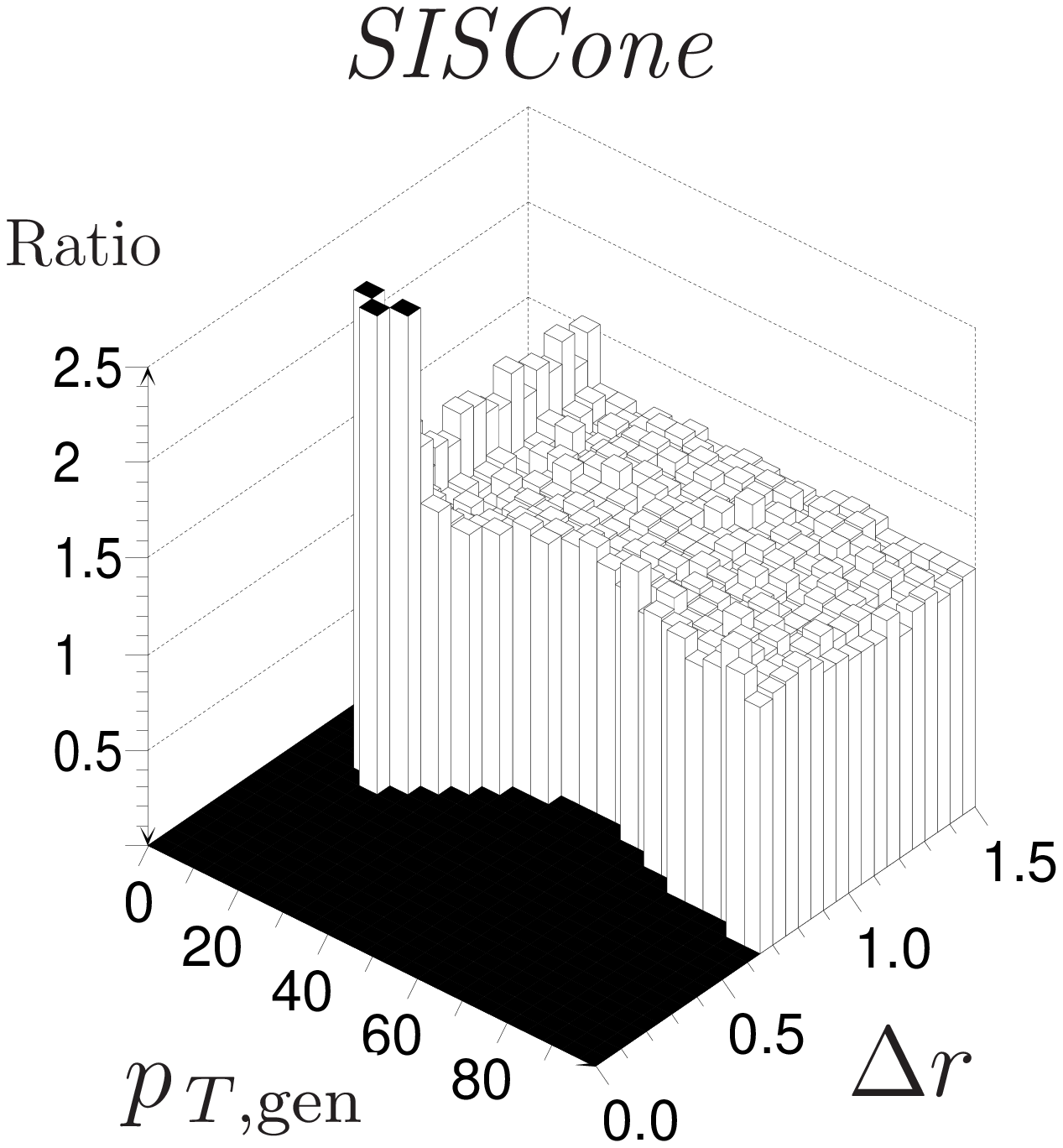} &
\epsfxsize=1.7in
\epsffile{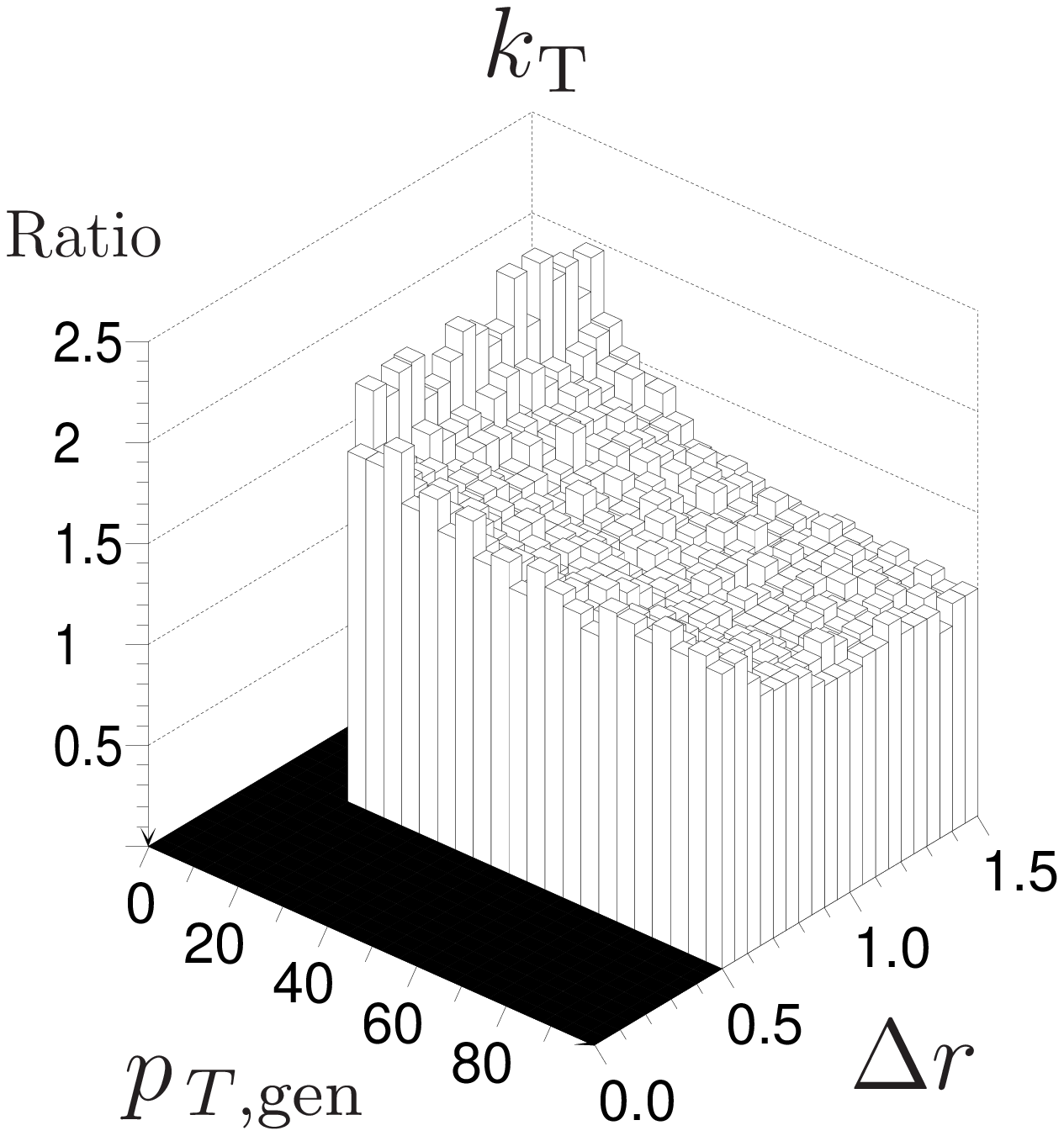} &
\epsfxsize=1.7in
\epsffile{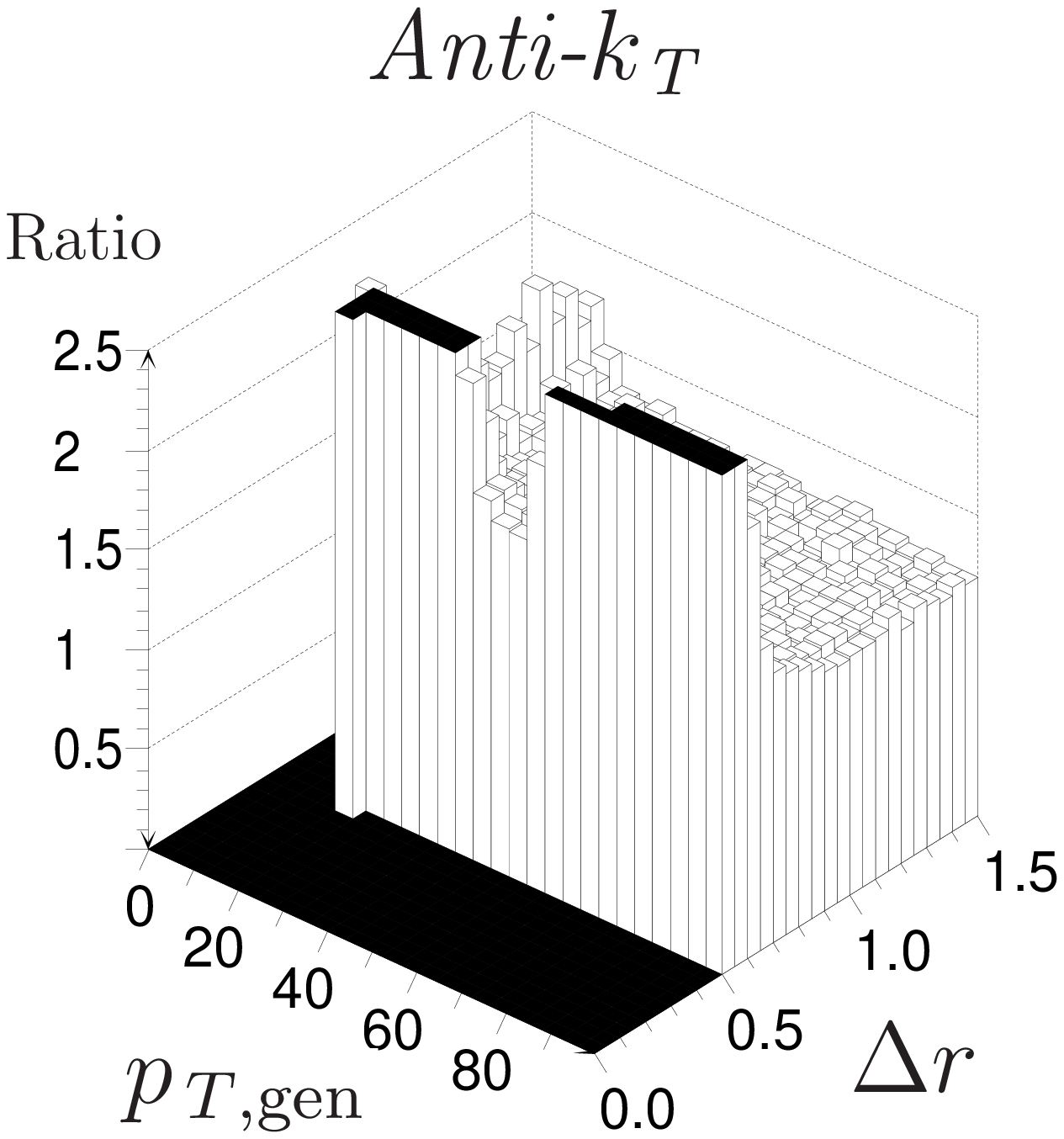} 
\end{array}$
\end{center}
\caption{$p_T$ reconstruction uncertainty ratios with {\it FFTJet-1} in the denominator.}
\label{fig:uncertrat}
\end{figure}

Figure~\ref{fig:uncertrat} presents the uncertainty ratios
in which the relative $p_T$ reconstruction uncertainties
for different algorithms are in the numerator
and the {\it FFTJet-1} uncertainty is in the denominator.
The division is performed bin-by-bin, and the result is set to zero
if either the numerator efficiency or the denominator
efficiency for that bin is less than 50\%.
Median uncertainty ratios (using non-zero bins only)
are listed in Table~\ref{table:recoresults}.
The table also shows the average fake rate which is
defined for the purpose of this study as the fraction of events which
have a reconstructed jet with $p_T > 10$~GeV/$c$
not matched to a generated jet ({\it i.e.}, there are two other
reconstructed jets which produce better matches).
\begin{table}[t!]
\caption{Summary of the results within the $\Delta r < 1.5$ and $p\sub{$T$,gen} < 100$~GeV/$c$ limits. These numbers
can be viewed as an~approximate guide for ordering jet algorithms 
according to their performance in a~strong magnetic field.}
\label{table:recoresults}
\begin{center}
\noindent\begin{tabular}{ccccc} \hline
Quantity & {\it FFTJet-1} & {\it SISCone} & \KT & {\it Anti}-\KT \\ \hline\hline
Average efficiency, \% & 65 & 51 & 64 & 64 \\
Average fake rate, \% & 0.1 & 0.3 & 0.7 & 0.2 \\
Median $p_T$ uncertainty ratio & 1.00 & 1.29 & 1.27 & 1.32 \\ \hline
\end{tabular}
\end{center}
\end{table}

Due to a more appropriate jet shape model (elliptical instead of circular),
{\it FFTJet-1} outperforms all other algorithms used in this study by
a~significant margin. The intrinsic jet
$p_T$ resolution uncertainties of other algorithms are
larger than the {\it FFTJet-1} uncertainty by $\approx$30\%.
It is clear from Fig.~\ref{fig:eff} that the
{\it SISCone} algorithm performance is significantly hampered
by the split-merge stage which creates a~complicated
efficiency dependence on $\Delta r$ and $p\sub{$T$,gen}$
and results in a reduced efficiency overall.
Because of
this problem, {\it SISCone} can not be recommended
for reconstructing multijet, high occupancy events.
It can also be argued that, compared to \KT and {\it Anti}-\KT$\!\!$,
the efficiency pattern exhibited
by {\it FFTJet-1} can potentially be more useful. {\it FFTJet-1}
remains efficient at smaller $\Delta r$ values instead of
the low $p\sub{$T$,gen}$ region where reliable jet reconstruction
is prevented by poor jet $p_T$ resolution.

\section{Acknowledgements}

The author
thanks the Jet Algorithms Group of the CMS collaboration for comments
and discussions.
The development of the FFTJet package was supported in part
by the US Department of Energy grant DE-FG02-95ER40938.

\end{document}